\documentclass[12pt,pre,aps,preprint]{revtex4}
\usepackage{psfig}

\newcommand{\bfm}[1]{\mbox{\boldmath ${#1}$}}
\newcommand{\BGY}{\bfm\Psi}

\begin{document}

\title{Optimal Bounds on the Trapping Constant 
and  Permeability of Porous Media}

\author{   S. Torquato}
\email[Corresponding author: ]{torquato@princeton.edu}
\author{D. C. Pham}
\affiliation{Department of Chemistry and Princeton Materials Institute,
Princeton University, Princeton, NJ 08544}

\date{\today}

\begin{abstract}

We derive exact expressions for so-called ``void''  bounds on the  trapping 
constant $\gamma$ and fluid permeability $k$
for  coated-spheres and coated-cylinders models of porous 
media.  We find that in some cases the bounds are optimal, i.e., the
void bounds coincide with the corresponding exact
solutions of $\gamma$ and $k$ for these coated-inclusions models.
In these instances, exact expressions are obtained for the relevant length
scale that arises in the void bounds, which depends on a two-point correlation function
that characterizes the porous medium.
In contrast to bounds on the effective
conductivity and elastic moduli of composite
media, this is the first time that model microstructures
have been found that exactly realize bounds
on either the trapping constant or fluid permeability.

\end{abstract}
\maketitle

The study of the effective 
properties of heterogeneous materials, such as composite
and porous media, has a rich history \cite{Ma73,Ei06} and
is a continuing  source of theoretically challenging questions
\cite{Jo86,Fe87,To90e,Cherk00,Mi02,To02a,Sa03}.
Except for a few special microstructures \cite{Cherk00,Mi02,To02a},
exact predictions of the effective properties are not possible because they
depend on an infinite set of statistical correlations that
characterize the microstructure \cite{To02a}. Thus, apart from such exact solutions, 
rigorous estimates
of effective properties must take the form of inequalities, i.e.,
upper and lower bounds \cite{Ha62c,Be68,Cherk00,Mi02,To02a}. 
Bounds are useful because often one of the bounds can provide
a useful estimate of the property even when the reciprocal
bound diverges from it \cite{To02a}. Moreover, 
it is highly desirable to determine optimal bounds and the microstructures
that attain them.  The best known bounds in the
cases of the effective conductivity and  bulk modulus
of two-phase media are the Hashin-Shtrikman bounds \cite{Ha62c}. These are optimal
bounds, given the phase volume fractions,
because they are realizable by, among other geometries, certain
coated-spheres and coated-cylinders assemblages in three
and two dimensions, respectively. 

Two important effective properties of fluid-saturated porous
media that have been extensively studied are the trapping constant
$\gamma$ \cite{Wi91,To91f,To02a} and scalar fluid permeability $k$ 
\cite{Sc74,Jo86,Sh88,Av91b,To02a,Sa03}. Bounds on  $\gamma$ \cite{Pr63b,Do76,Ta87,Ru88,To89a,To02a}
and  $k$ \cite{Pr61,Be85a,To87a,Ru89a,To02a} have been derived and computed. However, to
date, microstructures that exactly realize (or attain) any of these bounds
have yet to be identified. Torquato \cite{To02a} has observed that the difficulty 
in identifying optimal microstructures for these classes of problems lies in the 
fact that $\gamma$ and $k$ (unlike the conductivity and elastic moduli) are 
length-scale dependent properties and known bounds on them depend nontrivially on 
the specific forms of two-point and higher-order correlation functions. For example, 
the so-called {\it void} bounds on $\gamma$ and $k$ \cite{To89a,Ru89a,To02a} depend
on the two-point correlation function
$S_2(r)$ (defined below) and have been evaluated for various particle models
for the trapping constant \cite{To89a,To02a,Ma03}
and fluid permeability \cite{Pr61,Be85a,Ru89a,To02a}.

In this Letter, we  exactly evaluate the {\it void} bounds on the
trapping constant  $\gamma$ and fluid permeability $k$
for the coated-spheres and coated-cylinders 
models of porous media. Interestingly, we show that in
some cases the void bounds are optimal because they coincide with the corresponding exact
solutions of $\gamma$ and $k$ for these particular
coated-inclusions porous-media models. In these cases, 
we obtain exact expressions for the relevant length
scale that arises in the void bounds, which depends on $S_2(r)$.

Each realization of the porous medium occupies a region of $d$-dimensional space 
${\cal V}$ of volume $V$ that is partitioned into two disjoint 
regions: a pore space (phase) ${\cal V}_P$ of porosity $\phi_P$ and a 
solid space (phase) ${\cal V}_S$ of volume fraction $\phi_S=1-\phi_P$. 
Let $\partial{\cal V}$ denote the surface of interface 
between ${\cal V}_P$ and ${\cal V}_S$. The pore-space {\it indicator 
function} ${\cal I}^{(P)}({\bf x})$ is given by
\begin{equation}
{\cal I}^{(P)}({\bf x})= \left\{ \begin{array}{ll}
 1, & \mbox{${\bf x}$ in ${\cal V}_P$}\\
 0, & \mbox{otherwise}
\end{array}
\right.
\label{char-sphere}
\end{equation}
The indicator function ${\cal M}({\bf x})$ for the interface is defined as
\begin{equation}
{\cal M}({\bf x}) =\mid\nabla {\cal I}^{(P)}({\bf x})\mid\; .
\label{inter}
\end{equation}
For statistically homogeneous media, the ensemble averages of the indicator
functions (\ref{char-sphere}) and (\ref{inter}) are respectively equal to the 
phase volume fraction $\phi_P$ and  the specific surface $s$ (interfacial area per unit volume), i.e.,
\begin{equation}
\phi_{P}=\langle{\cal I}^{(P)}({\bf x})\rangle,  \qquad s=\langle{\cal M}({\bf x})\rangle  \; ,
\label{1-point}
\end{equation}
where angular brackets denote an ensemble average.

Before evaluating the bounds, we first define the effective
properties $\gamma$ and $k$, and the coated-spheres model.
First, consider the steady-state trapping problem \cite{To02a}. 
The reactant diffuses in the pore space ${\cal V}_P$ (i.e. trap-free region)
with scalar diffusion coefficient $D$
but is instantly absorbed when it makes contact 
with the interface between  ${\cal V}_P$ and the trap region ${\cal V}_S$.
At steady-state, the rate of production of the reactant $G$ (per unit trap-free volume)
is exactly compensated by its removal by the traps.
Two-scale homogenization theory \cite{Ru88}
enables one to show that the trapping constant $\gamma$ 
for a statistically homogeneous and ergodic medium obeys 
the first-order rate equation
\begin{eqnarray}
G=\gamma DC\; .
\end{eqnarray}
Here $C$ is the average concentration field and 
\begin{eqnarray}
\gamma^{-1}=\langle u\rangle =\lim_{V \rightarrow
\infty}\frac{1}{V}\int\limits_{\cal V} u({\bf x})d{\bf x}\; ,
\label{def-gamma}
\end{eqnarray}
where $u({\bf x})$ is the scaled concentration field
that solves the boundary-value  problem
\begin{eqnarray}
\nabla^2u({\bf x})&=&-1\; ,\quad {\bf x}\in {\cal V}_P\;
,\label{diff-gamma}\\
u({\bf x})&=&0\; ,\quad {\bf x}\in\partial{\cal V}\; \label{bc-gamma}.
\end{eqnarray}
It follows that the trapping constant $\gamma$ for any $d$ has
dimensions of the inverse of length squared \cite{To02a}.

Rubinstein and Torquato \cite{Ru88} have formulated a variational principle
in terms of the trial function
$v({\bf x})$ that enables one to obtain 
the following lower bound on $\gamma$ for ergodic
media:
\begin{equation}
\gamma \ge  \frac{1}
{\langle \nabla v({\bf x})\cdot \nabla v({\bf x}){\cal I}^{(P)}({\bf x})\rangle},
\label{var-gamma}
\end{equation}
where $v({\bf x})$ is required to satisfy the Poisson equation
\begin{equation}
\nabla^2v({\bf x})=-1\; ,\qquad{\bf x} \in {\cal V}_P\;.
\end{equation}
Elsewhere Torquato and Rubinstein \cite{To89a} constructed 
what they referred to as the {\em void} lower
bound in three dimensions by using a specific trial field.
The generalization of this trial field 
to any space dimension $d \ge 2$ \cite{To02a} is given by
\begin{equation}
v({\bf x})=\frac{1}{\phi_S}\int\limits_{\cal V} g({\bf x}-{\bf y})[{\cal
I}^{(P)}({\bf y})
-\phi_P]d{\bf y},
\label{trial-gamma}
\end{equation}
where
\begin{equation}
g({\bf r})= \left\{ \begin{array}{ll}
 \frac{\displaystyle 1}{\displaystyle 2\pi}\ln(\frac{\displaystyle 1}{\displaystyle r}), & \qquad d=2\\
 \frac {\displaystyle 1}{\displaystyle (d-2)\Omega(d)}\frac{\displaystyle 1}
{\displaystyle r^{d-2}}, & \qquad d\ge 3
\end{array}
\right.
\label{Green}
\end{equation}
is the $d$-dimensional Green's function for the Laplace operator 
\cite{To02a},
$\Omega(d)=2\pi^{d/2}/\Gamma(d/2)$ is the total solid angle
contained in a $d$-dimensional sphere,
$\phi_S$ is volume fraction of the trap phase, and $r \equiv |\bf r|$.
Substitution of  trial field (\ref{trial-gamma}) into the variational
principle (\ref{var-gamma}) yields the two-point 
void lower bound on $\gamma$ 
for general statistically homogeneous and isotropic $d$-dimensional porous
media \cite{To02a} as
\begin{equation}
\gamma \ge \frac{\phi_S^2}{\ell_P^2},
\label{void-gamma-S2}
\end{equation}
where $\ell_P$ is  a pore length scale defined by 
\begin{equation}
\ell_P^2 = \left\{ \begin{array}{ll}
 -{\int_{0}^{\infty}[S_2(r)-\phi^2_P] r\ln r dr}, & \qquad d=2\\ \\
\frac{\displaystyle 1}{\displaystyle (d-2)}{\int_{0}^{\infty}[S_2(r)-\phi^2_P] r  dr}      
, & \qquad d\ge 3,
\end{array}
\right.
\label{ell}
\end{equation}
and  $S_2({\bf r})$ is the two-point correlation function
defined by
\begin{equation}
S_2({\bf r})=\langle\cal I^{(P)}({\bf x})\cal I^{(P)}({\bf x+r})
\rangle\;.
\label{2-point}
\end{equation}
The function $S_2(\bf r)$ can also be interpreted as being the
probability of finding two points separated by the displacement
vector $\bf r$ in the pore space \cite{To02a}.

Using homogenization theory, Rubinstein and Torquato \cite{Ru89a} derived the conditions
under which the slow flow of an  incompressible viscous fluid 
through macroscopically 
anisotropic random porous medium is described by Darcy's law
\begin{eqnarray}
{\bf U}=-\frac{1}{\mu} {\bf k} \cdot \nabla p_0\; ,
\end{eqnarray}
where ${\bf U}$ is the average fluid velocity, $\nabla p_0$ is the 
applied pressure gradient, $\mu$ is the dynamic viscosity, and $\bf k$ is 
the symmetric fluid permeability tensor. 
In particular, for the special case of macroscropically 
isotropic media, the scalar fluid permeability
$k=\mbox{Tr}({\bf k})/d$ (where $\mbox{Tr}$
denotes the trace operation) is given by
\begin{equation}
k=\langle{\bf w} \cdot {\bf e}\rangle =\lim_{V \rightarrow \infty}\frac{1}{V}\int_
{\cal V} {\bf w}({\bf x})d{\bf x}\;,
\label{def-k}
\end{equation}
where $\bf e$ is a unit vector, and ${\bf w}$ and $\pi$ are, respectively,
a scaled velocity and scaled pressure that
satisfy the scaled Stokes equations 
\begin{eqnarray}
\nabla^2 {\bf w}&=&\nabla \pi- {\bf e}\quad\textrm{in}\quad {\cal V}_P\; ,\\
\nabla\cdot {\bf w}&=&0\quad\textrm{in}\quad {\cal V}_P\; ,\\
{\bf w}&=&{\bf 0}\quad\textrm{on}\quad\partial{\cal V}\;.
\end{eqnarray}
 It follows that the permeability
 $k$ for any $d$ has
dimensions of length squared \cite{To02a}.

Prager \cite{Pr61} was the first to derive a two-point ``void'' upper bound
on the permeability using a variational principle. Subsequently, 
Berryman and Milton \cite{Be85a}
corrected a normalization constraint in the Prager variational
principle using a volume-average approach. Rubinstein and Torquato\cite{Ru89a}
formulated new upper and lower bound variational principles employing an 
ensemble-average approach and also derived the void upper bound. 

For our purposes,
the Rubinstein-Torquato variational principle for the upper bound
is the most natural. It states that for ergodic media the  trial function
${\bf q}({\bf x})$ enables one to obtain 
the following upper bound on $k$:
\begin{equation}
k \le \langle {\bf q}({\bf x}): \nabla {\bf q}({\bf x}){\cal I}^{(P)}({\bf x})\rangle,
\label{var-k}
\end{equation}
where $\bf q({\bf x})$ is required to satisfy the momentum equation
\begin{equation}
\nabla\times\nabla^2 ({\bf q}+{\bf e})={\bf 0}\; ,
\qquad{\bf x} \in {\cal V}_P\; 
\end{equation}
Rubinstein and Torquato \cite{Ru89a} constructed the {\em void} upper
bound in three dimensions by using a specific trial field. The generalization
of this trial field to any dimension $d\ge 3$ is given by \cite{To02a}
\begin{eqnarray}
{\bf q}({\bf x})=\frac{1}{\phi_S}\int\limits_{\cal V}{\bf\BGY}({\bf x}-{\bf y})
\cdot {\bf e}[{\cal I}^{(P)}({\bf y})-\phi_P]d{\bf y},
\label{trial-k}
\end{eqnarray}
where 
\begin{equation}
{\BGY}({\bf r})= 
\frac{\displaystyle d}{\displaystyle (d^2-3)
\Omega(d) r^{d-2}}[{\bf I}+{\bf n}{\bf n}], \qquad d\ge 3,
\label{Green2}
\end{equation}
is the $d$-dimensional Green's function {\bf\BGY} (second order tensor) 
associated with the velocity for Stokes flow, ${\bf n}={\bf r}/r$,
and $\phi_S$ is the volume fraction of the obstacles. 
Substitution of  trial field (\ref{trial-k}) into the variational
principle (\ref{var-k}) yields the two-point void upper bound on $k$ 
for general statistically homogeneous and isotropic $d$-dimensional porous
media \cite{To02a} as
\begin{equation}
k \le \frac{(d+1)(d-2)}{d^2-3}\frac{\ell_P^2}{\phi_S^2}, \qquad d \ge 3,
\label{void-k-S2}
\end{equation}
where  $\ell_P$ is the length scale defined by  (\ref{ell}), which 
is precisely the 
same as the one that arises in the void lower bound on the 
trapping constant $\gamma$ for $d \ge 3$ \cite{footnote}.

\begin{figure}
\centerline{\psfig{file=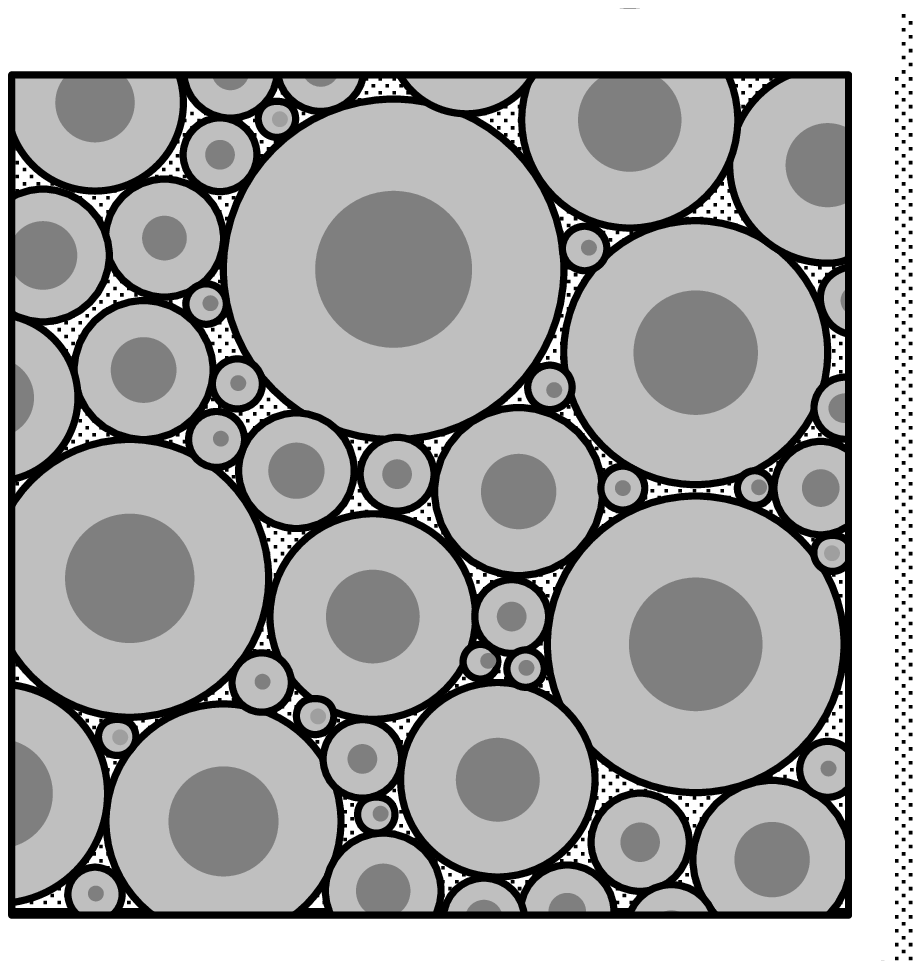,width=2.6in}}
\caption{Schematic of the coated-spheres model microstructure.}
\label{coated}
\end{figure}

The coated-spheres model \cite{Ha62c} consists of composite 
spheres that are composed of a spherical core of phase 2 (inclusion) and 
radius $R_I$, surrounded by a concentric shell of phase 1 (matrix) and 
outer radius $R_M$. The ratio $(R_I/R_M)^d$ is fixed and equal to the 
inclusion volume fraction $\phi_2$ in space dimension $d$. The composite 
spheres fill all space, implying that there is a distribution in their 
sizes ranging to the infinitesimally small (see Fig. \ref{coated}). The inclusion 
phase is always disconnected  
and the matrix phase is always connected (except at the trivial point 
$\phi_2=1$).

The coated-spheres model places restrictions on the size distribution
of the composite spheres.  Without loss of generality,
we will assume that the composite spheres
possess an infinite number of discrete sizes. 
Let $\rho_k$ be the number density (number of particles per unit volume) 
of the $k$th type of composite sphere of radius $R_{M_k}$ and let
$R_{I_k}$ denote the corresponding radius of the inclusion.
Moreover, we know that fraction of space covered by the composite
spheres, denoted by $\Phi$, is unity, and therefore we have
the following condition on the size distribution:
\begin{equation}
\Phi=\sum_{k=1}^{\infty} \rho_k v_1(R_{M_k})=1,
\label{Phi}
\end{equation}
where $v_1(r)$ is the $d$-dimensional volume of a single sphere of radius
$r$ given by 
\begin{equation}
v_1(r)=\frac{\pi^{d/2}}{\Gamma(1+d/2)}r^d.
\label{v1}
\end{equation}
For the volume fraction $\Phi$ to remain
bounded (i.e., for the sum (\ref{Phi}) to converge), 
$\rho_k R_{M_k}^d$ must also remain bounded
for all $k$. Thus,  the number
density $\rho_k$ must diverge to infinity
as $R_{M_k}$ approaches zero and 
the specific surface $s$ must also diverge, since 
$\rho_k R_{M_k}^{d-1}$ diverges as $R_{M_k}$ approaches zero.
Note that volume fraction $\phi_2$ of the inclusion phase
is given by
\begin{equation}
\phi_2=\sum_{k=1}^{\infty} \rho_k
v_1(R_{I_k})=\frac{\pi^{d/2}}{\Gamma(1+d/2)}\sum_{k=1}^{\infty}
\rho_kR^d_{I_k}.
\end{equation}
It is convenient to introduce the following $n$th moment of $R_I$:
\begin{equation}
\langle R^n_I \rangle =\frac{1}{\rho} \sum_{k=1}^{\infty} \rho_k R^n_{I_k},
\label{avg}
\end{equation}
where $\rho$ is a characteristic density
(e.g., inverse of the volume of the largest
composite sphere) and $n$ is any integer $n \ge 3$.
The inclusion volume fraction $\phi_2$ can now be 
reexpressed as
\begin{equation}
\phi_2 = \rho \frac{\pi^{d/2}}{\Gamma(1+d/2)} \langle R^d_I \rangle.
\label{phi}
\end{equation}

We first evaluate the void lower bound on $\gamma$ for the three-dimensional 
coated-spheres model. To begin,  we take the connected matrix phase ${\cal V}_1$ 
to be the traps and the disconnected  inclusion phase  ${\cal V}_2$ to
be the pore space. Therefore, the porosity is given by  $\phi_P=\phi_2$. Using the void
trial field (\ref{trial-gamma}) for $v(\bf x)$, we can obtain from 
(\ref{var-gamma}) the following lower bound on $\gamma$:
\begin{eqnarray}
\gamma\ge\Bigg[\lim_{V\rightarrow \infty }\frac{1}{V}\int\limits_{\cal V}\nabla v\cdot \nabla v{\cal I}^{(2)}({\bf x})d{\bf x}\Bigg]^{-1},
\label{lower1}
\end{eqnarray}
where we have equated ensemble averages with volume averages
via the ergodic hypothesis.
The key volume integrals can be evaluated following Pham \cite{Ph97}, 
and such details will be given elsewhere \cite{Ph04}. We find
that the void lower bound is exactly given by
\begin{eqnarray}
\gamma\ge \frac {15\langle R_I^3\rangle}{\phi_P\langle R_I^5\rangle}\; .
\label{void1}
\end{eqnarray}
Interestingly, by comparing this result to the general expression for the void 
upper bound (\ref{void-gamma-S2}), which is given in terms of the two-point 
correlation function $S_2(r)$, we see that the square of the pore length scale 
$\ell_P$ for $d=3$ is exactly given by
\begin{equation}
\ell_P^2=\int_{0}^{\infty}[S_2(r)-\phi^2_P] r  dr=
\frac{\phi_P\phi_S^2}{15} \frac{\langle R_I^5\rangle}{\langle R_I^3\rangle}.
\label{ell-coated3}
\end{equation}
for the coated-spheres model.

Now we show that  bound (\ref{void1}) coincides with 
the exact solution for this particular coated-spheres
model. Specifically, the exact solution of the boundary-value
problem (\ref{diff-gamma}) and (\ref{bc-gamma}) for diffusion inside a spherical 
inclusion ${\cal S}_I$ of radius $R_I$ is given by \cite{To02a} 
$u= (R_I^2-r^2)/6$ for $0 \le r \le R_I$. Hence, using definition
(\ref{def-gamma}), we find that $\gamma$, for {\it nonoverlapping}
sphere models with a {\it general size distribution} (not just the coated-spheres model)
is exactly given by
\begin{equation}
\gamma=\langle u\rangle^{-1} =\Bigg[\lim_{V \rightarrow
\infty}\frac{1}{V}\sum\limits_{{\cal S}_I\in {\cal
V}_2}\int_0^{R_I}\frac 16(R_I^2-r^2)4\pi r^2dr\Bigg]^{-1}=
\frac {15 \langle R_I^3\rangle}{\phi_P \langle R_I^5\rangle}.
\label{exact1}
\end{equation}

Thus, we see that the void lower bound (\ref{void1}) 
coincides with the exact solution (\ref{exact1}) for the coated-spheres model, 
and hence the bound is exactly realizable when the inclusions are taken to be 
the pore phase. This may immediately lead one to conclude 
that the void bound is optimal among all microstructures, but such
a statement cannot be made unless one attaches special conditions.
Recall that unlike the effective conductivity or effective elastic moduli,
the trapping constant as well as the fluid permeability are length-scale
dependent quantities. Thus, any statement about optimality must 
fix not only the porosity but the relevant length scales. The correct statement is the following:
The void bound is optimal among all microstructures that share
the same porosity $\phi_P$ and pore length scale $\ell_P$ as the
coated-spheres model [cf. Eq. (\ref{ell-coated3})].
One can always adjust the pore length scale (\ref{ell-coated3}) of the coated-spheres
model at some porosity $\phi_P$ to be equal to $\ell_P$ for 
any microstructure with same porosity.

As noted above, relation (\ref{exact1})
applies to diffusion within nonoverlapping
spheres with a general size distribution. Accordingly, let 
us define another squared length scale  
$L_P^2=\langle R_I^5\rangle /\langle 
R_I^3\rangle$ for such a general nonoverlapping sphere model.
In what follows, superscripts $g$ and $c$ are appended to quantities
associated with the general sphere model and coated-spheres model, respectively. 
The use of expressions (\ref{void-gamma-S2}), (\ref{void1}), and (\ref{exact1}) reveals the following
interrelations between these two models: at fixed $\phi_P$, if ${L}_P^{(g)}={L}_P^{(c)}$,
 then $\gamma^{(g)}=\gamma^{(c)}$ and $\ell_P^{(g)}\ge \ell_P^{(c)}$, and if 
$\ell_P^{(g)}=\ell_P^{(c)}$,  then $\gamma^{(g)}\ge\gamma^{(c)}$
and ${L}_P^{(g)}\le {L}_P^{(c)}$.

Next we take the connected matrix phase ${\cal V}_1$ to be the pore
space and the disconnected  inclusion phase  ${\cal V}_2$ to
be the traps so that $\phi_P=\phi_1$. Employing the 
trial field (\ref{trial-gamma}) for $v(\bf x)$, we  obtain the void lower bound for the model
as
\begin{eqnarray}
\frac{\gamma}{\gamma_s}\ge(1+\frac 15\phi_S^{1/3}-\phi_S-\frac 15\phi_S^2)^{-1}\; ,
\label{void2}
\end{eqnarray}
where $\gamma_s=3\phi_S\langle R_I^3\rangle/\langle
R_I^5\rangle$
and $\phi_S$ is the volume fraction of the traps. The exact solution in
this case is not known.

The procedure above can be repeated 
the two-dimensional coated-cylinders model. 
For diffusion inside circular inclusions, we obtain the void lower bound as
\begin{eqnarray}
\gamma\ge\frac {8\langle R_I^3\rangle}{\phi_P\langle R_I^5\rangle}\; ,
\end{eqnarray}
which coincides with the exact result, and thus is an optimal bound in the sense described above.
Comparison of this result to the general relation
for the void upper bound (\ref{void-gamma-S2}) yields
the following exact  expression for the square 
of the pore length scale $\ell_P$ for the
coated-cylinders model:
\begin{equation}
\ell_P^2=-\int_{0}^{\infty}[S_2(r)-\phi^2_P] r \ln r dr=
\frac{\phi_P\phi_S^2}{8} \frac{\langle R_I^5\rangle}{\langle R_I^3\rangle}.
\label{ell-coated2}
\end{equation}
For diffusion exterior to the circular inclusions, 
we obtain the void lower bound as
\begin{eqnarray}
\frac{\gamma}{\gamma_s} \ge 
(-\textrm{ln}\phi_S-\frac 32+2\phi_S-\frac 12\phi_S^2)^{-1}\;  
\label{void3}
\end{eqnarray}
where $\gamma_s=4 \langle R_I^3 \rangle\phi_S/\langle
R_I^5\rangle$.

Consider fluid flow along (inside or outside) bundles of parallel cylindrical 
circular tubes corresponding to the coated cylinders model. 
The velocity field reduces to an axial component only, and the Stokes equation reduces 
to a simple Poisson equation identical to that of the 2$D$ trapping problem. Hence, we have exactly the same solution for the axial component
of velocity as for the concentration field in the trapping problem,
leading to the exact result that $k=\gamma^{-1}$
\cite{To02a}. Exploiting this observation and using the previous 
results, we simply summarize the 
appropriate results below for $k$.

In particular, for axial flow inside the 
cylindrical tubes (Poiseuille flow), we obtain the void upper 
bound as
\begin{eqnarray}
k\le\frac {\phi_P\langle R_I^5\rangle}{8\langle R_I^3\rangle}\; ,
\end{eqnarray}
which coincides with the exact result \cite{To02a}, and thus is an optimal bound.
Comparison of this result to the general relation
for the upper bound (\ref{void-k-S2}) also yields
the exact formula (\ref{ell-coated2}) for the  
square of the pore length scale $\ell_P$ for the
coated-cylinders model in the {\it transverse} plane.

A well-known empirical estimate for $k$ is the Kozeny-Carmen 
relation \cite{To02a}
\begin{eqnarray}
k=\frac {\phi_P^3}{cs^2}\; ,
\end{eqnarray}
where $c$ is an adjustable parameter and $s$ is the specific surface
defined by (\ref{1-point}). However 
for the coated-spheres or coated-cylinders model with the inclusions of all sizes down to 
infinitesimally small, we saw earlier that $s$ diverges to infinity
and therefore the   Kozeny-Carmen relation incorrectly predicts
a vanishing permeability. This serves to illustrate the well-established  fact
that the permeability cannot generally be represented by
a simple length scale, such as the specific surface \cite{Jo86,Av91b,To02a}.

For flow exterior to the cylindrical  tubes, we obtain the void upper bound as
\begin{eqnarray}
\frac{k}{k_s}\le -\textrm{ln}\phi_S-\frac 32+2\phi_S-\frac 12\phi_S^2\; 
\label{void4}
\end{eqnarray}
where $k_s=\langle R_I^5\rangle/(4 \phi_s \langle
R_I^3\rangle)$.
For flow exterior to spherical obstacles, we exploit the fact that the void upper 
bound (\ref{void-k-S2}) on $k$ is trivially related to the void lower bound 
(\ref{void-gamma-S2}). Thus, we deduce the upper bound on $k$ for flow exterior
to spherical inclusions in the coated-spheres model
from the corresponding bound (\ref{void2}) on the trapping constant:
\begin{eqnarray}
k/k_s\le 1+\frac 15\phi_S^{1/3}-\phi_S-\frac 15\phi_S^2\;,
\label{void5}
\end{eqnarray}
where $k_s= 2\langle R_I^5\rangle /(9\phi_S\langle
R_I^3\rangle)$.

To summarize, the two-point void lower bound (\ref{void-gamma-S2}) on the trapping constant $\gamma$ and the void
upper bound (\ref{void-k-S2})  on the  fluid
permeability $k$ both generally depend on the pore length scale $\ell_P$,
defined by (\ref{ell}), which involves an integral over the two-point correlation function
$S_2(r)$ that characterizes the porous medium.
We have derived exact expressions for the void bounds on $\gamma$ and $k$ 
for certain coated-spheres and coated-cylinders models of porous media.
For diffusion inside the spherical ($d=3$) and cylindrical inclusions ($d=2$),
the void lower bound on $\gamma$ was shown to be exact. Similarly, for axial
flow inside the cylinders of the coated-cylinders model, 
the void upper bound on $k$ was demonstrated to be exact. In these instances, 
the void bounds are optimal among all microstructures that share
the space porosity $\phi_P$ and pore length scale $\ell_P$ as the
coated-spheres model. In contrast to bounds on the effective
conductivity and elastic moduli of composite
media, this is the first time that model microstructures
have been found that exactly realize bounds
on either the trapping constant or fluid permeability. 
For cases of diffusion and flow exterior to the spheres
and cylinders in the
coated-inclusions model of porous media,  exact results 
are not available, but we still obtained simple analytical expressions 
for the void bounds on $\gamma$ and $k$.
In future studies, it will be of interest to investigate
what are the optimal microstructures that correspond to
the improved two-point ``interfacial-surface" bounds
on both $\gamma$ and $k$, which also incorporate surface
correlation functions \cite{To02a}.

\begin{acknowledgments}
The work was completed during D. C. P.'s visit to the Materials Institute, 
Princeton University, as a Fulbright Senior Scholar. 
S. T.  was supported by the 
MRSEC Grant at Princeton University, NSF DMR - 0213706, and by the Air Force
Office of Scientific Research under Grant No.
F49620-03-1-0406.

\end{acknowledgments}


\begin{thebibliography}{10}
\bibitem{footnote}
The similarity between the void upper bound on $k$ and
the void lower bound on $\gamma$ indicates a deeper
relationship between these two properties, even 
though the corresponding governing field equations
are different, both in form and tensorial order. Indeed,
it was shown in Ref. 5 that the inverse of the isotropic trapping
constant tensor bounds the fluid permeability tensor
from above for general statistically homogeneous
porous media.
\end{thebibliography}

\end{document}